# Nonlinear modification of the laser noise power spectrum induced by a frequency-shifted optical feedback

Eric LACOT[1*], Bahram HOUCHMANDZADEH[2], Vadim GIRARDEAU[1], Olivier HUGON[1], and Olivier JACQUIN[1]

[1] *Univ. Grenoble Alpes, LIPhy, F-38000 Grenoble, France*

[2] *CNRS, LIPhy, F-38000 Grenoble, France*

[*] *Corresponding author: eric.lacot@ujf-grenoble.fr*

Abstract: In this article, we study the non-linear coupling between the stationary (i.e. the beating modulation signal) and transient (i.e. the laser quantum noise) dynamics of a laser subjected to frequency shifted optical feedback. We show how the noise power spectrum and more specifically the relaxation oscillation frequency of the laser are modified under different optical feedback condition. Specifically we study the influence of (i) the amount of light returning to the laser cavity and (ii) the initial detuning between the frequency shift and intrinsic relaxation frequency. The present work shows how the relaxation frequency is related to the strength of the beating signal and the shape of the noise power spectrum gives an image of the Transfer Modulation Function (i.e. of the amplification gain) of the nonlinear-laser dynamics.The theoretical predictions, confirmed by numerical resolutions, are in good agreements with the experimental data.

## I. Introduction

Laser properties (power, polarization, coherence, dynamical behavior …) can be significantly affected and modified by optical feedback [1,2]. These properties have been widely studied in the particular case of weak optical feedback. For example, it has been shown that the steady state intensity of a laser subjected to coherent optical feedback from the reflection on an external surface depends on the reflectivity, distance and motion of the surface. This led to the Laser Feedback Interferometry (LFI) technique [3]. It also has been demonstrated that the dynamical behavior of a laser can be several order of magnitude more sensitive to optical feedback than its steady state properties. Since the pioneering work of K. Otsuka on self-mixing modulation effect in a class-B laser [4], the high dynamical sensitivity of lasers to frequency shifted optical feedback has been used in metrology [5], for example in self-mixing Laser Doppler Velocimetry (LDV) [6-8] and in Laser Optical Feedback Imaging (LOFI) [9-11]. Compared to conventional optical heterodyne detection, frequency shifted optical feedback allows for several order of magnitude higher intensity modulation contrast [12-14]. In the case of weak optical feedback, the laser dynamics is linear and the maximum of the modulation is obtained when the frequency shift is resonant with the laser relaxation oscillation frequency. In this condition, an optical feedback level as low as -170 dB (i.e. $10^{17}$ times weaker than the laser intra-cavity power) has been detected [6]. When the optical feedback becomes stronger, nonlinearities appear in the laser dynamics [15, 16]. They can cause the apparition of chaos, bi-stability and hysteresis phenomenon with the tuning (back and forth) of the frequency shift of the optical feedback [17].

The main objective of the present work, is to study the modification of the noise power spectrum of the laser and more specifically its resonance frequency (i.e. the laser relaxation frequency) induced by the nonlinear laser dynamics in the strong feedback situation. A better understanding of the laser nonlinearbehavior could lead to a new generation of laser metrology techniques with improved performances.This article is organized as follows. In section II and III, we recall the rate equations governing the dynamics of a laser submitted to frequency shifted optical feedback.These equations are solved numerically to show characteristic examples of the modification of the laser noise power spectrum under different optical feedback conditions. In particular, the influence of (i) the amount of light returning to the laser cavity and (ii) the detuning between the frequency shift and intrinsic relaxation frequency are studied. Section III is devoted to analytical resolution of the rate equation: using a bifurcation analysis, we determine the amplitude of the laser output power modulation in the strong feedback situation. We show then how the nonlinear dynamical coupling links the value of the relaxation frequency to the strength of theoutput power modulation. Section IV is devoted to experimental results where we show their good agreement with the theoretical predictions of the preceding sections. The final section is devoted to the general discussion of these results and to their prospective applications.

## II. Laser with frequency shifted optical feedback

### A. Basic equations

For weak optical feedback ($R_e \ll 1$) and a short round trip time delay ($\tau_e \ll 1/F_e$), the dynamical behavior of a laser with frequency shifted ($F_e$) optical feedback can be described by the following set of differential equations [12]:

$$\frac{dI}{dt} = BIN - \gamma_c I + \gamma_c 2\sqrt{R_e} I \cos(\Omega_e t + \Phi_e) + F_I(t)$$

(1a)

$$\frac{dN}{dt} = \gamma_1 [N_0 - N] - BNI$$

(1b)

$$\langle F_I(t)\rangle = 0 \text{ et } \langle F_I(t)F_I(t-\tau)\rangle = 2\gamma_c\langle I\rangle\delta(\tau) \quad (1c)$$

where $I$ and $N$ are respectively the laser intensity (photon unit) and the population inversion (atom unit). $\gamma_1$ is the decay rate of the population inversion, $\gamma_c$ is the laser cavity decay rate, $\gamma_1 N_0$ is the pumping rate and $B$ is related to the Einstein coefficient (i.e. the laser cross section).

In Eq.(1a), the cosine function expresses the coherent interaction (i.e. the beating at the angular frequency: $\Omega_e = 2\pi F_e$) between the lasing and the feedback electric field. The optical feedback is characterized by the effective power reflectivity $R_e$ and the optical phase shift $\Phi_e = \omega_c \tau_e$ induced by the round trip time between the laser and the target (where $\omega_c$ is the optical pulsation). Regarding the noise, the laser quantum fluctuations are described by the Langevin noise function $F_I(t)$, with a zero mean value and a white noise type correlation function [Eq. (1c)] [18,19].

In the absence of optical feedback ($R_e = 0$) and noise ($F_I(t)=0$), the steady-state of Eqs. (1a-c) is given by:

$$N_S = \gamma_c/B \quad (2a)$$

$$I_S = I_{sat}[\eta-1] \quad (2b)$$

where $\eta = N_0/N_S$ is the normalized pumping parameter and $I_{sat} = \gamma_1/B$ is related to the saturation intensity of the laser transition.

In this regime, the intrinsic dynamics of a class-B laser ($\gamma_c \gg \gamma_1\eta$) is characterized by damped relaxation oscillations of the laser output power with a relaxation angular frequency $\Omega_R = \sqrt{\gamma_1\gamma_c(\eta-1)}$ and a damping rate $\Gamma_R = \gamma_1\eta/2$. Experimentally, this transient dynamics is constantly excited by the laser quantum noise described by the Langevin force $F_I(t)$.

In the presence of a strong optical feedback, the laser intrinsic dynamics and in particular the relaxation oscillation frequency is modified and will depend on the modulation conditions ($R_e, \Omega_e$). In this article, we call "*strong feedback*", the regime where the modulation frequency is nearly resonant ($\Omega_e \approx \Omega_R$) and where the amount of optical feedback ($R_e$) is high enough to induce non-linear dynamical behaviors (which give rise to the generation of harmonic and parametric dynamical frequencies) in the laser output power modulation [13, 15, 16]. In contrast, the "*weak feedback*" regime corresponds to the situation where the modulation frequency is far away from the resonance ($|\Omega_e - \Omega_R| \gg 0$) and where the amount of optical feedback ($R_e$) is small enough to induce only linear dynamical behaviors in the laser output power modulation.

## B. Numerical results

Using a standard Runge-Kutta method, we have solveda normalized form of Eqs. (1a-c), where the laser intensity is divided by the saturation intensity ($i = I/I_{sat}$), the population inversion is divided by its stationary value ($n = N/N_S$) and the time is multiplied by the cavity damping rate ($t' = \gamma_c t$). The normalized Langevin force is a random variable with zero mean value and a standard deviation equal to: $1/\sqrt{I_{sat} dt'}$ where the integration step is: $dt' = 40$. The number of integration steps is equal to: $16384$.

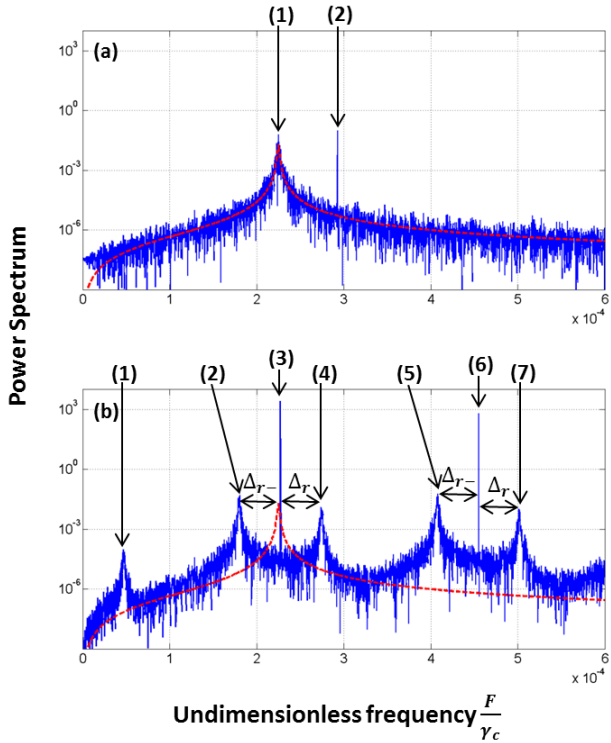

Fig.1: Numerical simulation. Power spectra of the laser intensity dynamics $|FT[I(t)/I_{sat}]|^2$. a) Weak feedback: $R_e = 10^{-11}$, $F_e/F_R = 1.3$; Peaks: (1) $F_R$, (2) $F_e$. b) Strong feedback: $R_e = 10^{-7}$, $F_e/F_R = 1.01$; Peaks: (1) $F_e - \tilde{F}_R$, (2) $\tilde{F}_R$, (3) $F_e$, (4) $2F_e - \tilde{F}_R$, (5) $F_e + \tilde{F}_R$, (6) $2F_e$, (7) $3F_e - \tilde{F}_R$. Laser parameters: $I_{sat} = 3 \times 10^9$ photons, $\eta = 1.2$, $\gamma_c = 1 \times 10^9$ $s^{-1}$, $\gamma_1/\gamma_c = 1 \times 10^{-5}$, $F_R/\gamma_c = 2.251 \times 10^{-4}$. The dashed curve is a fit of the laser noise power spectrum without feedback.

Fig. 1 shows the RF power spectra of the laser output power modulation for two different optical feedback conditions ($R_e, \Omega_e$). When the feedback is weak (Fig. 1a), one can observe that the laser dynamics is principally composed of the superposition of the laser output power modulation at the modulation frequency $F_e$ and of the noise power spectrum related to the transient relaxation oscillations, with a resonant frequency $F_R = \Omega_R/2\pi$ and a half width at half maximum $\Gamma_R/2\pi$. In the strong feedback regime (Fig. 1b), the power spectrum is composed of peaks at the modulation frequency ($F_e$) and its harmonics ($2F_e,...$) while the noise generates two sidebands beside each peak, located at $F_e \pm \Delta_R$, $2F_e \pm \Delta_R,...$. In this study, we define the new value of the relaxation frequency $\tilde{F}_R$ as the position of the highest noise peak in the vicinity of $F_e$. In Fig. 1b, it corresponds to the left sideband: $\tilde{F}_R = F_e - \Delta_R$ (2 in Fig. 1b). One can notice that all other peaks can be obtained from a linear combination of $F_e$ and $\tilde{F}_R$. For example, the right sideband of $F_e$ corresponds to: $F_e + \Delta_R = 2F_e - \tilde{F}_R$ (4 in Fig. 1b). The comparison of Figs. 1a and 1b shows how the noise power spectrum is modified by the strong optical feedback at the modulation frequency ($F_e$) and how the new relaxation frequency ($\tilde{F}_R$) is moved down from its intrinsic position ($F_R$).

By comparison with the work made in [15,16], which is focused on the study of the signal power spectrum, the present work analyzes how the noise power spectrum of the laser is modified and how the relaxation frequency of the laser is shifted, when the laser is submitted to a strong optical feedback.

Figure 2 captures the general feature of laser noise power spectrum modification when the amount of optical feedback ($R_e$) increases and/or when the modulation frequency approaches the intrinsic laser relaxation frequency ($F_e \to F_R$).

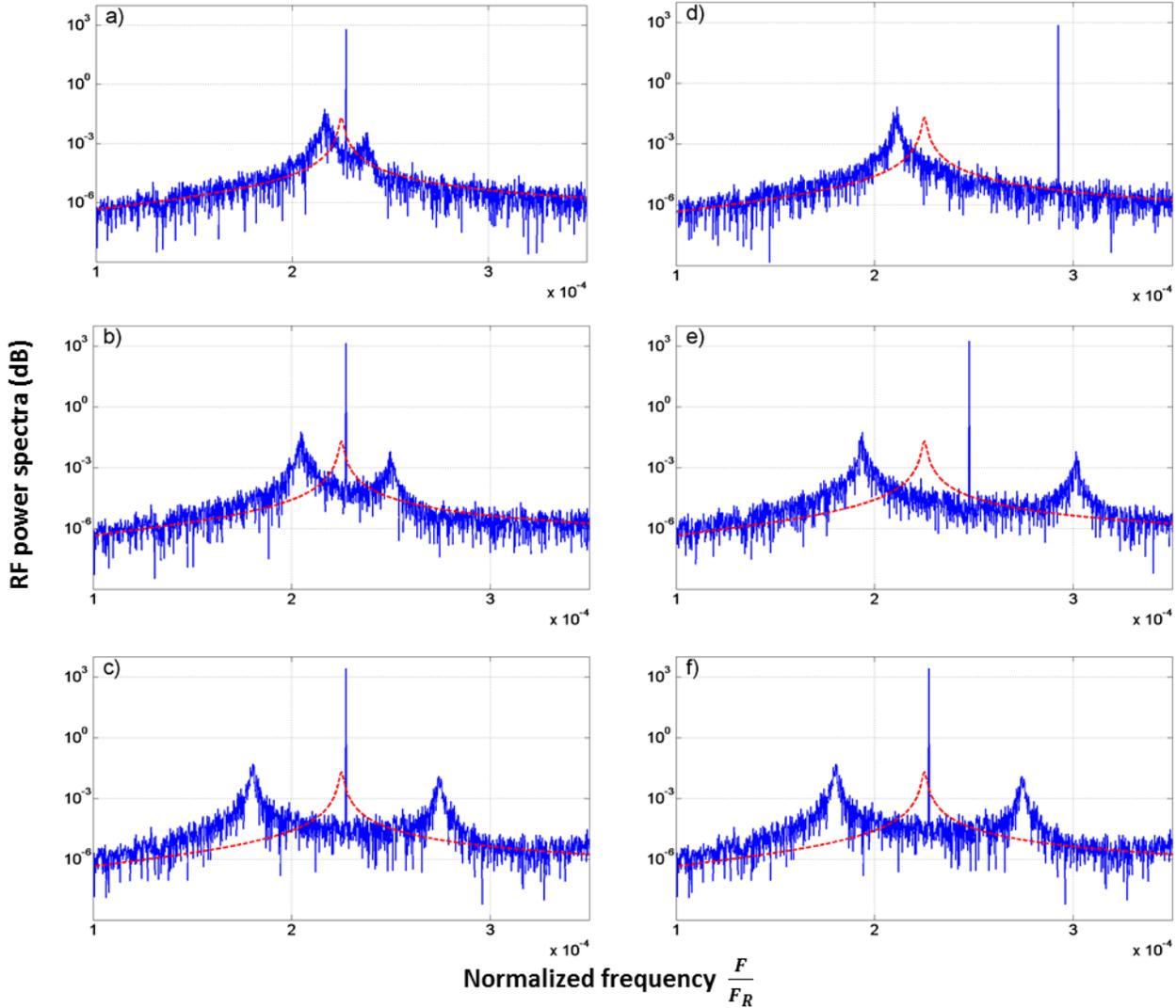

Fig.2: Numerical simulation. Power spectra of the laser intensity dynamics. Left column: fixed modulation frequency ($F_e/F_R = 1.01$) and an increasing amount of optical feedback ($R_e$); a) $R_e = 10^{-9}$; b) $R_e = 10^{-8}$, c) $R_e = 10^{-7}$. Right column: fixed amount of optical feedback ($R_e = 10^{-7}$) and a modulation frequency going closer to the intrinsic relaxation frequency; d) $F_e/F_R = 1.3$; e) $F_e/F_R = 1.1$; f) $F_e/F_R = 1.01$. Laserparameters are identical to Fig. 1. The dashed curve is a fit of the laser noise power spectrum without feedback.

For a given modulation frequency, the left column of Fig. 2 shows how $\tilde{F}_R$ (i.e. the noise left side-band) is shifted to lower frequencies when the amount of optical feedback increases. For a given amount of optical feedback, the right column of Fig. 2 shows how $\tilde{F}_R$ is pushed down to a lower frequency when the modulation frequency approaches the intrinsic laser relaxation frequency.

Also, one can observe that in the low feedback condition the noise power spectrum exhibits a single resonance (Figs. 1a and 2d) while in the in strong feedback condition it exhibits a double resonance (see Figs. 2a-c and 2e-f) due to a nonlinear dynamical coupling effect. Here, let us mentioned that the shape of the noise power spectrum is an image of the Modulation Transfer Function (i.e. of the amplification gain) of the laser dynamics.

## C. Bifurcation analysis

To study the dynamical response of a laser subject to frequency-shifted optical feedback ($R_e \neq 0$), we have made a bifurcation analysis using an asymptotic approximation similar to [17, 20, 21]. As already mentioned, we focus our analysis on the case $\Omega_e \approx \Omega_R$, where $\Omega_R$ is the intrinsic relaxation oscillation frequency of the laser. For a more convenient theoretical analysis, we have reformulated the set of Eqs. (1a-c) by using the new variables (x and y), the new time (s) , and the new modulation parameters ($\delta_e, \sigma_e$) defined by [17, 21]:

$$I = I_S(1+y), \quad N = N_S\left(1 + \frac{\Omega_R}{\gamma_c}x\right), \quad s = \Omega_R t \quad (3a)$$

$$\sigma_e = \frac{\Omega_e}{\Omega_R}, \quad \delta_e = 2\sqrt{R_e}\frac{\gamma_c}{\Omega_R} \quad (3b)$$

The new variables x and y are, respectively, the deviation from the nonzero intensity laser steady-state and the population inversion steady state. In the absence of noise, the set of Eqs. (1a-b) can then be rewritten as:

$$\frac{dx}{ds} = -y - \varepsilon x[1 + (\eta-1)(1+y)] \quad (4a)$$

$$\frac{dy}{ds} = (1+y)[x + \delta_e \cos(\sigma_e s + \Phi_e)] \quad (4b)$$

For our microchip laser, $\varepsilon = \gamma_1/\Omega_R \approx 10^{-3}$ is a small quantity which motivates an asymptotic analysis of Eqs. (4a-b). For $\sigma_e \approx 1$, the leading approximation of the solutions for the deviations x and y from the nonzero intensity state can be written as:

$$x = A\exp(iS) - i\frac{A^2}{3}\exp(i2S) + cc \quad (5a)$$

$$y = -iA\exp(iS) - \frac{2}{3}A^2\exp(i2S) + cc \quad (5b)$$

where $S = \sigma_e s = \Omega_e t$. The complex amplitude $A = 0(\varepsilon^{1/2})$ of the small periodic oscillation is a functions of the slow time $\varepsilon s$ and satisfies the following equations [17, 21]:

$$\frac{dA}{ds} = -i[\sigma_e - 1]A - i\frac{A^2 A^*}{6} - \frac{\eta \varepsilon}{2} A + i\frac{\delta_e}{4}\exp(i\Phi_e) \quad (6)$$

Notes that our asymptotic approximation is valid in the limit $\varepsilon \to 0$ assuming that $\sqrt{R_e}$ and $|\sigma_e - 1|$ are at maximum $0(\varepsilon)$ quantities.

In the right-hand side of Eqs. (5a-b), the first term corresponds to the laser output power modulation at the modulation frequency (i.e., at the cavity loss modulation frequency). The second term corresponds to the first harmonic modulation (i.e. non-linear laser dynamics) induced by the non-linear coupling between the laser intensity and the laser population inversion through the stimulated emission.

## III. Asymptotic solutions

We now investigate the solutions of Eq. (6), and show how the intrinsic laser dynamics (and in particular the laser relaxation frequency) can be influenced by the optical feedback through the modulation conditions $(\delta_e, \sigma_e)$. In this study, the key is the third order term $(-iA^2 A^*/6)$.

### A. Linear case ($A^2 A^* \approx 0$)

When the product $A^2 A^*$ can be neglected in Eq. (6), the solution of the amplitude equation can be written as small transient (with subscript T) relaxations of random excitations (due to quantum noise) around the stationary solution (with subscript S):

$$A(s) = A_{S,L} + B_{T,L}\exp(-\beta s) \quad (7a)$$

where:

$$A_{S,L} = \frac{i\delta_e \exp(i\Phi_e)}{4i(\sigma_e - 1) + 2\eta\varepsilon} \quad (7b)$$

$$\beta = i(\sigma_e - 1) + \eta\varepsilon/2 \quad (7c)$$

and the $B_{T,L}$ factor is given by the amplitude of the quantum noise.

By combining these results with Eqs. (3a-b) and (5a-b), we find that the stationary laser output power modulation is then given by:

$$I_S y_{S,L} = 2\sqrt{R_e} I_S |G_L(\Omega_e)|\sin(\Omega_e t + \arg(A_{S,L})) \quad (8a)$$

$$|G_L(\Omega_e)| = \left|\frac{\gamma_c(\Omega_e + \eta\gamma_1)}{\Omega_R^2 - \Omega_e^2 + i\Omega_e\eta\gamma_1}\right|$$

$$\approx \frac{\gamma_c}{\sqrt{4(\Omega_e - \Omega_R)^2 + (\eta\gamma_1)^2}} \quad (8b),$$

where $G_L(\Omega_e)$ is the amplification gain of the linear laser dynamics [13,14].

The above equation shows that the laser output power modulation exhibits a strong resonance (and therefore a high sensitivity to optical feedback) when $\Omega_e = \Omega_R$. The linear regime is valid when $|A_{S,L}| \ll 1$, which implies $R_e \ll R_{e,\lim}$ where: $R_{e,\lim} = \left(\frac{\gamma_1 \eta}{2\gamma_c}\right)^2$.

The transient laser output power modulation can be obtained from the relaxing part of Eq. (7):

$$I_S y_{T,L} = 2I_S |B_{T,L}| \exp\left(-\frac{\gamma_1 \eta}{2}t\right) \sin\left(\Omega_R t + \arg(B_{T,L})\right) \quad (9)$$

Experimentally, this transient dynamic is constantly excited by the laser quantum noise described by the Langevin forces ($F_I(t)$ in Eqs. (1a-c)). The RF power spectrum of the laser noise is obtained from the Fourier Transform (FT) of Eq. (9):

$$PS_L(\Omega) \propto |FT[I_S y_{T,L}(t)]|^2 \approx [I_S 2|B_{T,L}|/\gamma_c]^2 |G_L(\Omega)|^2 \quad (10)$$

In previous studies [12-14], we had already demonstrated that the detection of frequency shifted optical feedback is shot noise limited and that the RF power spectrum is given by :

$$PS_L(\Omega) \approx 2I_{out} |G_L(\Omega)|^2 2\Delta F \quad (11),$$

where $I_{out} = \gamma_c I_S$ is the laser output power and $\Delta F$ is the detection bandwidth which is supposed to be smaller than the resonance width ($\Delta F \ll \Gamma_R/2\pi$).

Eqs. (10) and (11) show that shape of the noise power spectrum is linked to the amplification gain of the laser dynamics ($G_L(\Omega)$) and that the amplitude of the intensity relaxation oscillations induced by the laser quantum noise is proportional to the square root of the laser output power :

$$I_S |B_{T,L}| \propto \sqrt{2I_{out} 2\Delta F} \quad (12)$$

## B. Non-linear case ($A^2 A^* \neq 0$)

In this section, we study the coupling of the stationary modulation of the laser output power with the laser transient dynamics (i.e. the laser quantum noise) through the non-linear cubic term $(-A^2 A^*/6)$ of Eq. (6). More precisely the aim of this study is to determine how the intrinsic laser dynamics (and in particular the laser relaxation frequency) is modified by the optical modulation conditions induced by the optical feedback. When the product $A^2 A^*$ can't be neglected in Eq. (6), the solution of the amplitude equation can be written as:

$$A(s) = A_{S,NL} + B_{T,NL}(s) \quad (13)$$

where $A_{S,NL}$ is the stationary solution and $B_{T,NL}(s)$ the transient dynamics of random excitations (due to quantum noise) around the stationary solution in the nonlinear regime (subscript NL).

In agreement with our numerical simulations (Figs. 1b and 2), we assume that the laser output power modulation is strongest than the laser quantum noise ($|A_{S,NL}| \gg |B_{T,NL}(s)|$). Keeping only the first order nonlinear terms in Eq. (13) gives:

$$A^2 A^* \approx A_{S,NL} A_{S,NL} A^*_{S,NL} + 2A_{S,NL} A^*_{S,NL} B_{T,NL} + A_{S,NL} A_{S,NL} B^*_{T,NL} \quad (14)$$

The insertion of Eqs. (13) and (14) into Eq. (6) gives the following equalities:

$$i\frac{\delta_e}{4}\exp(i\Phi_e) = i\left[\frac{\eta\varepsilon}{2}+\sigma_e-1\right]A_{S,NL} \\ +i\frac{A_{S,NL}^2 A_{S,NL}^*}{6} \quad (15a)$$

$$\frac{dB_{T,NL}}{ds} = -\left[\frac{\eta\varepsilon}{2}+i[\sigma_e-1]\right]B_{T,NL} \\ +i\frac{A_{S,NL}A_{S,NL}^*}{3}B_{T,NL} - i\frac{A_{S,NL}^2}{6}B_{T,NL}^* \quad (15b)$$

Eqs. (15a-b) clearly shows that the stationary and the transient dynamics (and therefore the laser noise) and are coupled through the nonlinear laser dynamics.

### 1. Nonlinear stationary solutions

Taking the complex conjugate of Eq. (15a) and multiplying them together, we find the square modulus $R_{S,NL}^2 = A_{S,NL}A_{S,NL}^*$ of the modulation amplitude:

$$\frac{R_{S,NL}^6}{36} + \frac{2(\sigma_e-1)}{6}R_{S,NL}^4 \\ +\left[(\sigma_e-1)^2+\left(\frac{\eta\varepsilon}{2}\right)^2\right]R_{S,NL}^2 - \left(\frac{\delta_e}{4}\right)^2 = 0 \quad (16)$$

which is a third order equation in $R_{S,NL}^2$. For a given amount of optical feedback $\delta_e$, the above equation has three real positive roots when the frequency shift is in a range defined by (see [17] or Appendix A for details):

$$-\left(\frac{\delta_e/(4\sqrt{6})}{\eta\varepsilon/2}\right)^2 < (\sigma_e-1) < -\frac{3}{2^{2/3}}\left(\frac{\delta_e}{4\sqrt{6}}\right)^{2/3} \quad (17)$$

The jump between the two stable solutions allows one to observe a hysteresis phenomenon which has already been studied in [17].

This kind of hysteresis has also been observed in the dynamics of a laser submitted to a an optical injection [22] or a modulation of the pumping power [23]. More specifically, the effect of noise on the size of the hysteresis zone has been studied in [24].

Let us recall that the goal of the present paper is not to study the hysteresis phenomenon, but to determine how the noise power spectrum of the laser and more specifically the laser relaxation frequency are related to the strength of $R_{S,NL}^2$.

### 2. Nonlinear relaxation oscillations

To determine the noise power spectrum we have studied the transient dynamics of random excitations ($B_{T,NL}$).

Writing $B_{T,NL}(s) = \exp\left(-\frac{\eta\varepsilon}{2}s\right)[X(s)+iY[s]]$, Eq. (15b) gives:

$$\frac{d}{ds}\begin{pmatrix}X\\Y\end{pmatrix} = \begin{pmatrix}\beta_Y & (\Gamma-\beta_X)\\-(\Gamma+\beta_X) & -\beta_Y\end{pmatrix}\begin{pmatrix}X\\Y\end{pmatrix} \quad (18)$$

Where $\beta_X+i\beta_Y = A_{S,NL}^2/6$, $\Gamma = (\sigma_e-1)+R_{S,NL}^2/3$ and $R_{S,NL}^2$ is the square modulus of the nonlinear laser output power modulation given by Eq. (16).

The two eigenvalues ($\lambda$) of the above matrix obey the equation:

$$\lambda^2 = \left(\frac{R_{S,NL}^2}{6}\right)^2 - \left((\sigma_e - 1) + \frac{R_{S,NL}^2}{3}\right)^2 \quad (19)$$

For the two stable solutions, one has $\lambda^2 < 0$, then $\lambda \in \mathbb{C}$ and $\lambda = \pm i\delta_{R,NL}$, where $\delta_{R,NL}$ allows one to measure the detuning of the transient frequency compared to the modulation frequency:

$$\delta_{R,NL} = \sigma_e - \tilde{\sigma}_R = \frac{\Omega_e}{\Omega_R} - \frac{\tilde{\Omega}_R}{\Omega_R}, \quad (20)$$

where $\tilde{\Omega}_R$ is the new value of the laser relaxation frequency modified by the nonlinear coupling between the transient and the permanent dynamics. Note that knowing $\sigma_e$ (which is experimentally controlled) and measuring the shift of the lateral side bands $\pm\delta_{R,NL} = \pm\frac{\Delta_R}{F_R}$ (see Fig.1b), one could determine the relative variation of the laser relaxation oscillation frequency:

$$\frac{\tilde{\Omega}_R - \Omega_R}{\Omega_R} = \sigma_e - 1 - \delta_{R,NL} \quad (21)$$

To make a simple physical analysis of this results, let us look at a particular situation where the optical feedback is not too strong ($R_{S,NL}^2 \ll (\sigma_e - 1)$). In this condition Eq. (19) simply gives:

$$\delta_{R,NL} \approx (\sigma_e - 1) + \frac{R_{S,NL}^2}{3}, \quad (22)$$

and using Eq. (19), one obtains the relative variation of the laser relaxation frequency:

$$\frac{\tilde{\Omega}_R - \Omega_R}{\Omega_R} \approx -\frac{R_{S,NL}^2}{3}. \quad (23)$$

Eq. (23) clearly shows that the frequency shift of the laser relaxation frequency is always negative and is proportional to the square of the amplitude of the laser output power modulation. Therefore the frequency shift is important for a strong optical feedback ($R_{S,NL}^2 \approx 1$), while it is roughly equal to zero for a weak optical feedback ($R_{S,NL}^2 \ll 1$). One can notice that due to the fact that $R_{S,NL}^2(\sigma_e)$ can exhibit hysteresis [17], $\tilde{\Omega}_R(\sigma_e)$ can also exhibits hysteresis.

Fig. 3 shows (i) the relative variation of the relaxation frequency (Fig. 3a) and (ii) the variation of the amplitude of the laser output power modulation (Fig. 3b) versus the detuning of the modulation frequency, for two different values of the optical feedback ($R_e = 1 \times 10^{-7}$, $R_e = 1 \times 10^{-8}$).

Fig. 3 shows a relatively good agreement between the analytical and the numerical results for both the amplitude of the laser output power modulation and the resonance frequency of the noise power spectrum (i.e. the laser relaxation frequency). In agreement with Eq. (23), one can observe that the decrease of the relaxation frequency is proportional to the increase of the laser amplitude modulation.

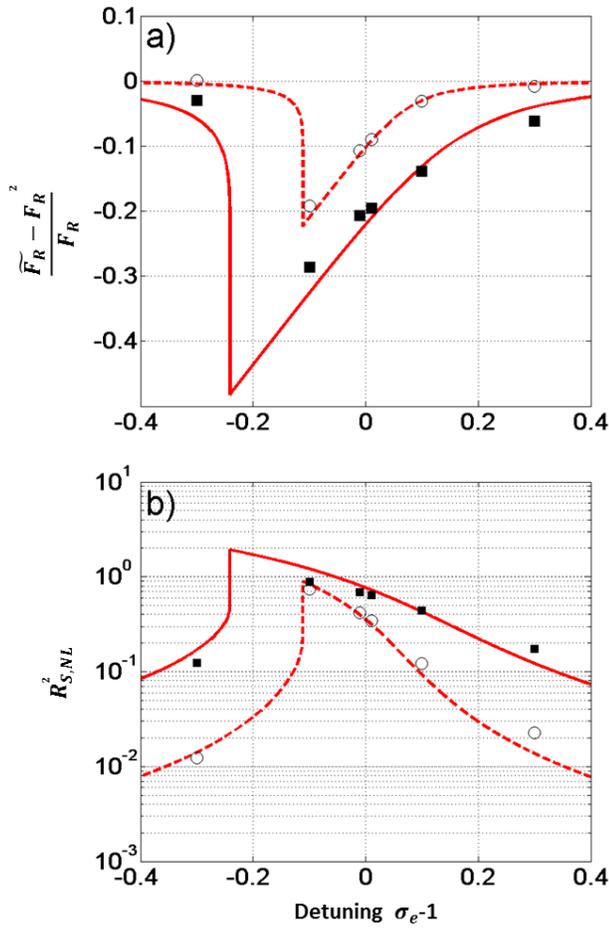

Fig.3 (Color online). (a): Relative variation of the relaxation frequency and (b): square of the amplitude of the laser output power modulation . The lines correspond to results obtained from analytical equations [Eq. (16)] while the dots correspond to results obtained from numerical simulations. Solid lines and square dots ($R_e = 1\times10^{-7}$), doted lines and circular dots ($R_e = 1\times10^{-8}$). Laser parameters are identical to Fig. 1.

On Fig.3, to avoid confusion, the hysteresis phenomenon is not shown (the frequency shift is not sweep back and forth). The square and the circle correspond simply to two different values of the optical feedback. Note also that the results shown on this figure have been obtained without any approximation or adjustment.

Finally, the RF power spectrum of the laser noise is obtained from the Fourier Transform (FT) of $B_{T,NL}$:

$$PS_{NL}(\Omega) \propto \left|FT\left[I_S y_{T,NL}(t)\right]\right|^2 \quad (24a)$$
$$\approx \left[I_S 2|B_{T,NL}(0)|/\gamma_c\right]^2 \left\langle|G_{NL}(\Omega)|\right\rangle^2$$

$$\left\langle|G_{NL}(\Omega)|\right\rangle \approx \left|\frac{\gamma_c(i\Omega+\gamma_1\eta)}{(\Omega_e - 2\pi\Delta_R)^2 - \Omega^2 + i\Omega\gamma_1\eta}\right|$$
$$+ g_{NL}\left|\frac{\gamma_c(i\Omega+\gamma_1\eta)}{(\Omega_e + 2\pi\Delta_R)^2 - \Omega^2 + i\Omega\gamma_1\eta}\right| \quad (24b)$$

$$g_{NL} \approx \left(\frac{2\frac{R^2_{S,NL}}{6}}{4\left[\sigma_e - 1 + 2\frac{R^2_{S,NL}}{6}\right]}\right)^{1/2} \quad (24c)$$

where $G_{NL}(\Omega)$ is the amplification gain (i.e. the modulation transfer function) of the nonlinear laser dynamics. This gain is composed of two resonance frequenciessymmetrically located of both side of the frequency shift: $(F_e \pm \Delta_R)$, where $g_{NL} \approx \frac{\left\langle|G_{NL}(\Omega_e - 2\pi\Delta_R)|\right\rangle}{\left\langle|G_{NL}(\Omega_e + 2\pi\Delta_R)|\right\rangle}$ is the ratio between the two maxima which depends on the feedback conditions.

In agreement with the numerical resultsshowed on Fig.2, Eqs. (24a-c) show that the amplitude of the two maxima of the noise power spectrum are of the same order of magnitude for strong optical feedback (i.e. when $R_{S,NL}^2 \approx 1$). Inversely, the right maximum disappears for weak optical feedback (i.e. when $R_{S,NL}^2 \ll 1$) leading to the conventional linear gain with only one resonance frequency ($\Omega_R$): $G_{NL}(\Omega)\underset{R_{S,NL}^2 \to 0}{\approx} G_L(\Omega)$.

One can notice an average $\langle \ \rangle$ in Eq. (24a). This average is due to the fact that $B_{T,NL}(s)$ is the transient dynamics of random excitations (due to quantum noise). With a uniform distribution of initial phase condition, this average leads to an incoherent interaction (modulus sum in Eq. (24b)) between the two resonance curves.

## IV. Experimental results

### A. Experimental setup

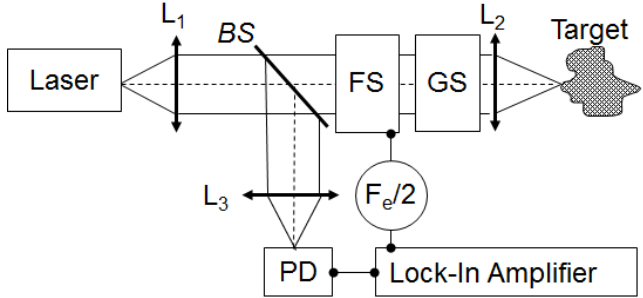

Fig. 4. Schematic diagram of the LOFI setup. L1, L2 and L3: Lenses, BS: Beam Splitter, GS: Galvanometric Scanner, FS Frequency Shifter with a round trip frequency-shift $F_e$, PD: Photodiode.

To study the nonlinear dynamics of a laser submitted to frequency shifted optical feedback, we have used a LOFI (Laser Optical Feedback Imaging) setup [9]. A schematic diagram of this setup is shown in Fig. 4. The laser is a diode pumped Nd:YAG microchip laser. The maximum available pump power is 400 mW at 810 nm and the total output power of the microchip laser is 80 mW with a central wavelength of $\lambda = 1064\,nm$. This laser has a plane-parallel cavity which is stabilized by the thermal lens induced by the Gaussian pump beam. The two dielectric mirrors are directly coated on the laser material (full cavity). The input dichroic mirror transmits the pump power and totally reflects the infrared laser wavelength. On the other side, the dichroic output mirror allows to totally reflect the pump power (to increase the pump power absorption and therefore the laser efficiency) and only partially reflects (95%) the laser wavelength. The microchip cavity is relatively short $L_c \approx 1mm$ which ensures a high damping rate of the cavity and therefore a good sensitivity to optical feedback. Part of the light diffracted and/or scattered by the target returns inside the laser cavity after a second pass through the frequency shifter. Therefore, the optical frequencies of the reinjected light are shifted by $F_e$. This frequency can be adjusted and is typically of the order of the laser relaxation frequency $F_R$, which is in the megahertz range for the microchip laser used in this study.

The optical feedback is characterized by the complex target reflectivity ($r_e = \sqrt{R_e}\exp(j\Phi_e)$, where the phase $\Phi_e = \frac{2\pi}{\lambda}d_e$ describes the optical phase shift induced by the round trip time delay $\tau_e$ (i.e. the distance $d_e = c\tau_e$, where $c$ is the velocity of light) between the laser and the target. The effective power reflectivity ($R_e = |r_e|^2$) takes into account the target albedo, the numerical aperture of the collection optics, the frequency shifters efficiencies, the transmission of all optical components and the overlap of the retro-diffused field with the Gaussian cavity beam (confocal feature).

The coherent interaction (beating) between the lasing electric fields and the frequency-shifted reinjected fields leads to a modulation of the laser output power at the frequency $F_e$. For detection purposes, a small part of the laser output beam is sent to a photodiode. The delivered voltage is analyzed by a numerical oscilloscope which allows Fast Fourier Transform (FFT) calculations, and processed by a lock-in amplifier which gives the LOFI signal (i.e. the amplitude and the phase of thelaser output power modulation) at the demodulation frequency $F_e$. Experimentally, LOFI images can be obtained pixel by pixel by a full 2D galvanometric scanning. In this study, the scanning device is not used because all the measures are realized on a single target point. One can notice that in contrast to a conventional heterodyne interferometer, the LOFI setup shown here does not require complex alignment. Indeed, the LOFI interferometer is always self-aligned because the laser simultaneously fulfils the functions of the source (i.e. photons-emitter) and of the photo-detector (i.e. photons-receptor).

## B. Experimental observations of the relaxation frequency shift

Fig. 5 shows, how the experimental noise power spectrum of the laser is modified when the amount of optical feedback ($R_e$) increases and/or when the modulation frequency approaches the intrinsic laser relaxation frequency ($F_e \approx F_R$).

In agreement with Eqs. (24a-c), Fig. 5 shows that the noise power spectrum gain is composed of two resonance frequencies symmetrically located of both side of the frequency shift: $(F_e \pm \Delta_R)$, and that the amplitude of the two maxima are of the same order of magnitude for strong optical feedback (i.e. when $R_{S,NL}^2 \geq -20 dB$ in arbitrary unit). Inversely, the right maximum disappears for weak optical feedback (i.e. when $R_{S,NL}^2 \leq -20 dB$) leading to the conventional linear gain with only one resonance frequency ($F_R$).

For a given modulation frequency, the left column of Fig.5 shows how $\tilde{F}_R$ (i.e. here the left side band) is shifted down to lower frequencies when the amount of optical feedback increases.

For a given amount of optical feedback, the right column of Fig. 5 shows how $\tilde{F}_R$ is shifted to a lower frequency when the modulation frequency decreases. Note that the numerical results of Fig. 2 and the experimental results of Fig. 5 are qualitatively similar.

More quantitatively, Fig. 6 shows the relative variation of the relaxation frequency (Fig. 6a),

and the variation of the amplitude of laser output power modulation (Fig. 6b) as a function of the modulation frequency. A good agreement between the analytical and the experimental results can be observed.

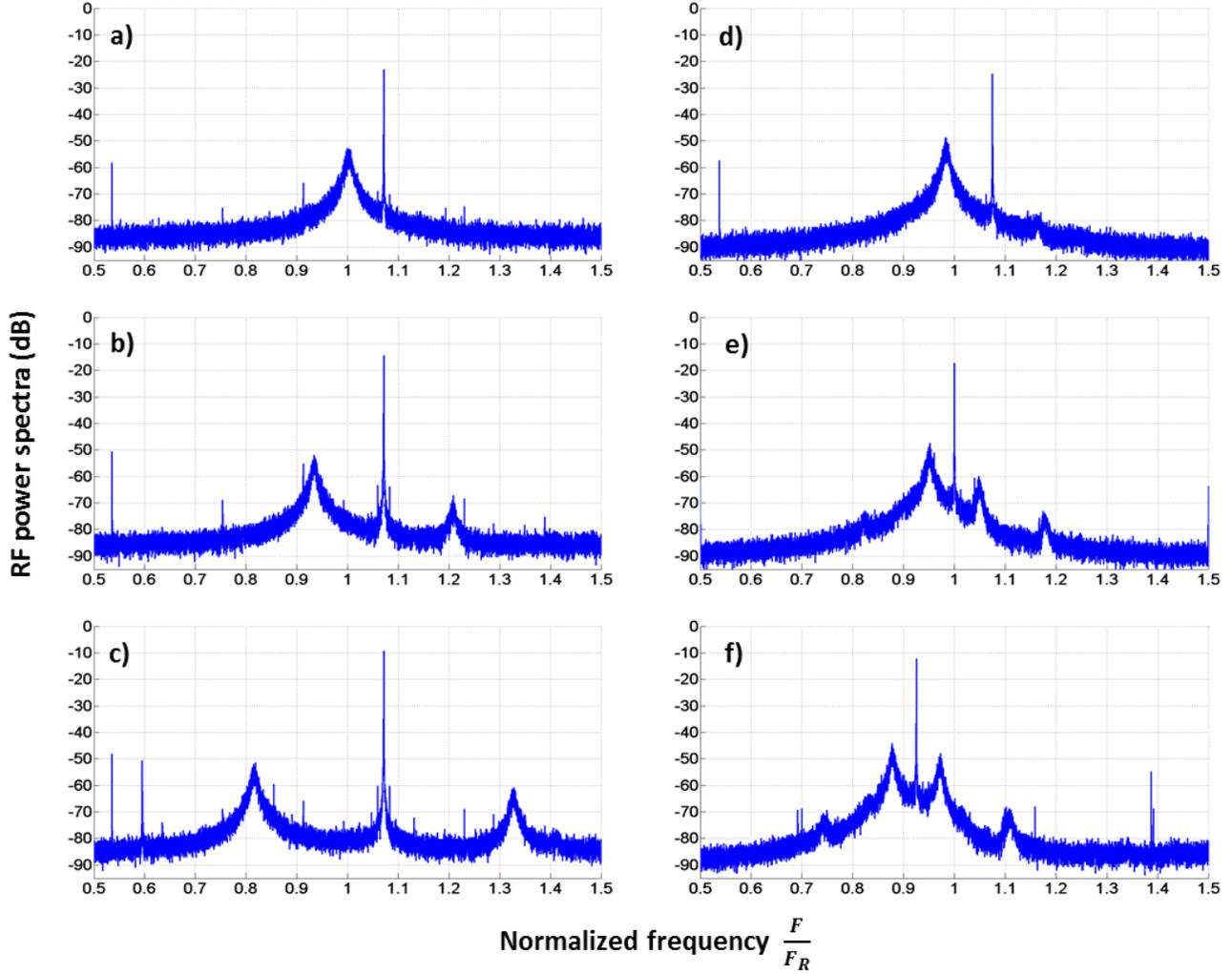

Fig.5: Experimental power spectra of the laser intensity dynamics. Left column, fixed modulation frequency ($F_e/F_R = 1.075$) and an increasing amount of optical feedback $R_e$ characterized by an increase of the laser output power modulation: a) $R^2_{S,NL} = -23.1 dB$; b) $R^2_{S,NL} = -14.3 dB$; c) $R^2_{S,NL} = -9.27 dB$. Right column, fixed amount of optical feedback and a decreasing modulation frequency around the intrinsic relaxation frequency: d) $F_e/F_R = 1.075$, $R^2_{S,NL} = -24.72 dB$; e) $F_e/F_R \approx 1.0$, $R^2_{S,NL} = -19.18 dB$; f) $F_e/F_R = 0.925$, $R^2_{S,NL} = -12.15 dB$.

First, one can observe in Figs. 6a and 6b that the amplitude of the modulation ($R^2_{S,NL}$) at the frequency shift ($F_e$) and the relaxation frequency shift $\tilde{F}_R$ as a function of the detuning vary in opposite direction. Indeed, in agreement with Eq. (23), the decrease of the relaxation frequency is proportional to the increase of the square of the laser amplitude modulation. Secondly, one can observe a small hysteresis induced by the direction (up or down) of the modulation frequency variation.

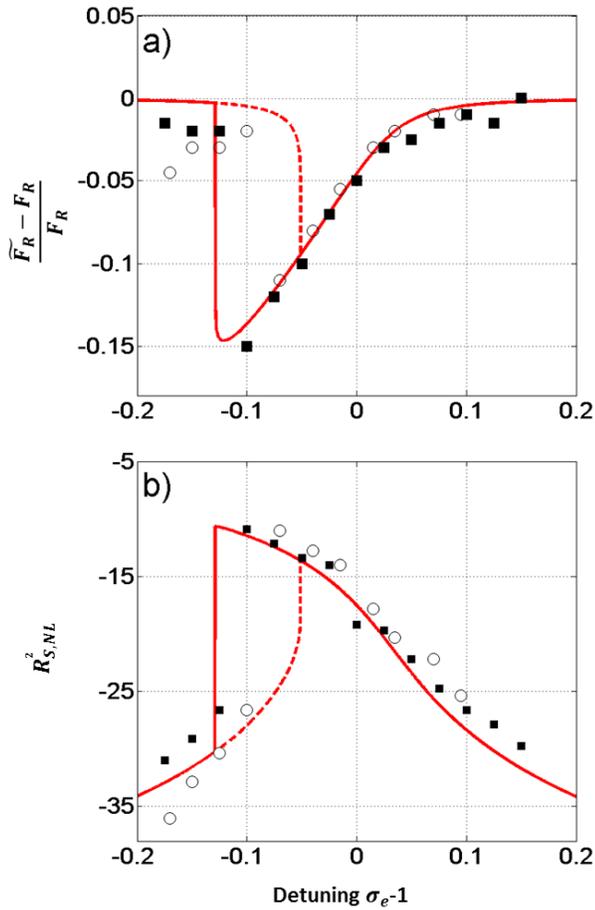

Fig. 6 (Color online). (a) Relative variation of the relaxation frequency and (b) square of the amplitude of the laser output power modulation versus the normalized scanning of the modulation frequency. The lines correspond to results obtained from analytical equations while the dots correspond to results obtained from experimental measurement. Dashed lines and circular dots: the scanning of the modulation frequency increases. Solid lines and square dots: the scanning of the modulation frequency decreases. Estimated laser parameters: $\eta = 1.2$, $\gamma_1 = 1\times 10^5\ s^{-1}$, $\gamma_c = 1\times 10^9\ s^{-1}$, $F_R \approx 700\ kHz$. Estimated feedback parameter: $R_e = 1\times 10^{-8}$.

In agreement with Eq. (23), one can understand that the hysteresis observed in the relaxation frequency shift is induced by the hysteresis of the modulation amplitude previously studied in [17]. The experimental results shown on Fig. 6 have been analytically adjusted by using a feedback reflectivity of $R_e \approx 1\times 10^{-8}$ and the following values of the laser parameters: $\eta = 1.2$, $\gamma_1 = 1\times 10^5\ s^{-1}$, $\gamma_c = 1\times 10^9\ s^{-1}$, corresponding to an intrinsic laser relaxation frequency of $F_R \approx 700\ kHz$.

## V. Conclusion and perspectives

In this paper we have demonstrated (analytically, numerically and experimentally) how the stationary dynamics (i.e. the output power modulation) and the transient dynamics (i.e. the laser quantum noise) of a laser subjected to frequency shifted optical feedback are coupled through the nonlinear laser dynamics. Both the numerical results and the experimental ones show a very good agreement with the analytical predictions. More precisely, this study shows how the noise power spectrum is related to the strength of the beating between the intra-cavity laser electric field and the frequency-shifted optical electric field (i.e. the amplitude of the laser output power modulation). The shape of the noise power spectrum gives an image of the Transfer Modulation Function (i.e. of the amplification gain) of the nonlinear-laser dynamics. A better understanding of the laser nonlinear behavior could lead to a new generation of laser metrology techniques with improved performances.

More specifically, this study shown how the relaxation frequency is shifted and how this shift becomes significant in the strong feedback situation where the modulation frequency is nearly resonant ($\Omega_e \approx \Omega_R$) and where the amount of optical feedback is high enough ($R_e > R_{e,\lim} = \left(\dfrac{\gamma_1 \eta}{2\gamma_c}\right)^2$) to induce non-linear dynamical behaviors in the laser output power modulation. Under these conditions, we have also observed that the relaxation frequency shift exhibits a small hysteresis, induced by the direction (up or down) of the variation of the modulation frequency. This hysteresis phenomenon appears simultaneously in the amplitude of the laser power modulation and in the shift of the relaxation oscillations frequency (i.e. resonant peak of the noise). This correlation demonstrates again the nonlinear dynamical coupling between these two physical quantities.

From the applied optics point of view, since the relaxation frequency, and consequently the noise power spectrum of the laser, can be controlled by the strength of the optical feedback, it opens interesting new perspectives in the field of interferometric phase measurement (and more particularly vibrometry) using a LOFI setup. For example, when the amount of optical feedback increases, the noise in the vicinity of the modulation frequency seems to be pushed far away, allowing lower phase noise and consequently the possibility to detect lower amplitudes of vibration [25,26]. Also, the shape of the noise power spectrum gives an image of the Transfer Modulation Function (i.e. of the amplification gain) of the nonlinear laser dynamics. The knowledge of this gain seems to be of importance to allow measurement of transient vibrations with sub-nanometric amplitude and a broad vibration spectrum of several megahertz (around the carrier frequency $F_e$). This work is in progress.

---

# APPENDIX A: Existence of three real roots for the amplitude of the laser output power modulation

The square modulus of the stationary modulation amplitude in the nonlinear case [Eq. (16)] obeys the equation:

$$R_{S,NL}^2 \left[ \left( \sigma_e - 1 + \frac{R_{S,NL}^2}{6} \right)^2 + \left( \frac{\eta \varepsilon}{2} \right)^2 \right] = \left( \frac{\delta_e}{4} \right)^2 \quad \text{(A1)}$$

Setting $R_{S,NL}^2/6 = \alpha x$, where $\alpha = \left(\delta_e/4\sqrt{6}\right)^{2/3}$ the above equation transforms into:

$$\left[(a+x)^2 + b^2\right] = \frac{1}{x} \quad (A2)$$

where $a = (\sigma_e - 1)/\alpha$ and $b = (\eta\varepsilon/2)/\alpha$. This equation is geometrically understood as the crossing of a parabola and a hyperbola (Figure A1).

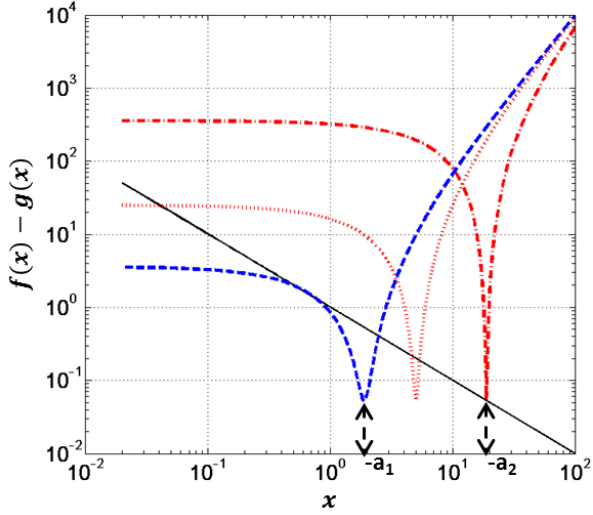

Fig. A1 (Color online). LogLog plot of the hyperbola $g(x) = 1/x$ (solid line) and three parabola $f(x) = (x+a)^2 + b^2$ (dash, dot and dash-dot lines) for a fixed value of $b = 0.23$ and three values of $a$ : ( $a_1 = -3/2^{2/3} \approx -1.89$, $a = -5$, $a_2 \approx -1/b^2 = -18.85$ ).

For a given value of $b$, there exists a range $a \in [a_1, a_2]$ where the equation has three real roots. To find the boundary of this range, we have to solve simultaneously Eq. (A2) and:

$$2(x+a) = -\frac{1}{x^2} \quad (A3)$$

where $b$ is given and $a$ and $x$ are to be determined. The exact solution is obtained by solving first for $x$:

$$\frac{1}{4x^4} - \frac{1}{x} + b^2 = 0 \quad (A4)$$

and then use Eq. (A3) to get the value of $a_i$.

For $b \ll 1$, the computation is simplified and the boundaries are approximately given by (Fig. A1):

$$a_1 = -3/2^{2/3}, \quad a_2 = -1/b^2 \quad (A5)$$

In terms of real parameters, the above condition is written:

$$-\left(\frac{\delta_e/(4\sqrt{6})}{\eta\varepsilon/2}\right)^2 < (\sigma_e - 1) < -\frac{3}{2^{2/3}}\left(\frac{\delta_e}{4\sqrt{6}}\right)^{2/3} \quad (A6)$$

Eq. (A6) is identical to Eq. (17).